# Do surfaces with mixed hydrophilic and hydrophobic areas enhance pool boiling?

**Amy Betz[1], Jie Xu[2], Huihe Qiu[3], Daniel Attinger[1*]**

[1]Mechanical Engineering, Columbia University, New York, NY .

[2]Mechanical Engineering, Washington State University, Vancouver, WA

[3]Mechanical Engineering, Honk Kong University of Science and Technology, Hong Kong

We demonstrate that smooth and flat surfaces combining hydrophilic and hydrophobic patterns improve pool boiling performance. Compared to a hydrophilic surface with 7º wetting angle, the measured critical heat flux and heat transfer coefficients of the enhanced surfaces are up to respectively 65 and 100% higher. Different networks combining hydrophilic and hydrophobic regions are characterized. While all tested networks enhance the heat transfer coefficient, large enhancements of critical heat flux are typically found for hydrophilic networks featuring hydrophobic islands. Hydrophilic networks indeed are shown to prevent the formation of an insulating vapor layer.

PACS: [insert here]

---

[*]*Corresponding author: da2203@columbia.edu*

Boiling is an efficient process to transfer large amounts of heat at a prescribed temperature because of the large latent heat of vaporization. The term *flow* boiling describes the boiling of liquids forced to move along hot surfaces, while in *pool* boiling, the topic handled in this paper, the liquid is stagnant and in contact with a hot solid surface [1]. Besides the common experience of boiling water in an electric kettle, pool boiling has applications in metallurgy, high performance heat exchangers, and immersion cooling of electronics. Pool boiling performance is measured with two parameters, the heat transfer coefficient (HTC) and the critical heat flux (CHF). The CHF is measured by increasing the surface temperature until a transition from high HTC to very low HTC occurs. This signifies the formation of a vapor film insulating the liquid from the heated surface, a phenomenon called dry out. Several characteristics determine the performance of a boiling surface. First, nucleation sites in appropriate number and dimensions need to be provided such as cavities, rough areas or hydrophobic islands [2]. As of today, the performance of boiling surfaces has been increased by using wicking structures to prevent dry out [3], by increasing the surface area with fins or fluidized bed [3-6], and by enhancing the wettability of the surface [5-9]. The latter strategy is justified by experiments of Wang and Dhir [10], showing that the CHF was increased by enhancing surface wettability. Wettability can be enhanced by either increasing the surface roughness or with microstructure or nanostructure coatings. For instance, Jones et al. [11] have shown that a well-chosen roughness can double or triple the HTC. Significant heat transfer enhancement has also been obtained with surfaces coated with a µm-thick carpet of nanometer diameter rods (nanorods) [5-7]. The CHF enhancement was attributed to coupled effects such as the multi-scale geometry [5,7] and the superhydrophilicity of the nanowire arrays [6,7].

A common assumption[1] in boiling studies is that the surface has a unique value of wettability. However the above introduction shows that the influence of wettability on boiling is complex: while hydrophobic zones promotes nucleation, the surface hydrophilicity does enhance the CHF [10]. In this work we take advantage of microlithography techniques to design surfaces combining hydrophobic and hydrophilic zones, for pool boiling experiments. Our intuition is that a well-designed network of hydrophobic and hydrophilic regions might promote nucleation, enhance the HTC, and even increase the CHF by preventing dry out.

Several patterns have been designed and fabricated using photolithography as shown in *Figure 1*, with hydrophobic and hydrophilic regions. The pattern size *d* was typically between 40 and 60% of the pitch p between patterns. Pattern shapes were hexagonal. We varied the pitch p from 50 to 200 μm, as well as the connectivity of the hydrophobic and hydrophilic patterns. Hydrophilic surfaces with hydrophobic islands were called *hydrophilic networks,* and noted *(+)*, meaning that any two hydrophilic regions could be joined without passing over a hydrophobic zone. Hydrophobic surfaces with hydrophilic islands were called *hydrophobic networks,* and noted *(-)*. The patterns were produced with a photolithographic process. First, the oxidized two-sided polished silicon wafer is coated with a 25nm Aluminum layer. Then, a hydrophobic coating of Teflon (AF400, Dupont), diluted in Fluorinert (FC-40) at a 1:3 ratio, is spun on top of the Aluminum layer to a thickness of about 100nm. After baking at 90°C for 20 minutes, 1μm layer of positive photoresist (S1818, Shipley) is spun on top of the Teflon. The wafer is then exposed to 180 mJ/cm$^2$ of UV light using a transparency mask. The exposed area and the underlying Teflon and Aluminum are then removed using oxygen plasma and a developer (300 MIF, AZ Electronic Materials). Finally, the remaining positive photoresist is removed with acetone. Another manufacturing process using self-assembled monolayers of octadecyltrichlorosilane (OTS) on $SiO_2$ [12] was tested but abandoned because of delamination during the boiling experiments. On the bottom side of the wafer an Indium Tin Oxide (ITO) heater is sputtered, with thermally evaporated Copper electrodes. The heater is then passivated with a 100nm $SiO_2$ layer. A thin thermocouple (Omega CO2, K-type) is taped onto the center of the ITO heater using a polyimide adhesive pad. A 5mm thick PDMS layer is then used to seal the bottom side of the wafer. Optionally, a final step was added to increase the wettability, by rinsing the top side of the wafers for a few seconds with a diluted solution of buffered hydrofluoric acid (HF). The wettability was measured with a goniometer, with wetting angles of 110º for the Teflon, 10-25º for the $SiO_2$ and 7º for the $SiO_2$ treated with HF. The maximum height of the hydrophobic patterns was 100 nm, while the roughness was below 5nm, as measured with a contact profilometer (Dektak). No difference in pattern height was observed after the HF treatment.

Heat transfer measurements are run using a pool boiling setup similar to the one in [6]. The wafer is placed in a polycarbonate chamber open to the atmosphere, filled with degassed, deionized water. The water is maintained at the saturation temperature of 100ºC by two immersed 100W cartridge heaters. A 750W power supply (Agilent N5750A) is used in constant voltage mode to apply a given heat flux to the 1 cm² ITO heater patterned on the back of the wafer. A data acquisition system (OMEGA DAQ-3000) is used to record the temperature measured on the back of the wafer, $T_{meas}$. The temperature at the wafer-water interface $T_w=T_{meas}-q''t/k$ is then determined using Fourier's law, with $q''$, $t$ and $k$ the respective heat flux, wafer thickness and Silicon thermal conductivity. For each data point presented in Figure 2, the temperature is obtained by averaging three hundred readings over about three minutes. The CHF is determined as the heat flux corresponding to the last observed stable temperature, beyond which a sudden dramatic increase in temperature is observed. The maximum combined uncertainty on the heat flux was estimated as ±1.5 W/cm², caused by the measurement of the heater area and the measurement of the electrical power. The maximum uncertainty on the superheat was estimated as ±1.5 K, due to the thermocouple uncertainty, temperature acquisition and heater/wafer thickness measurement uncertainties. For superheat values larger than 1 K, the uncertainty on the HTC is typically ±3 kW/m²K.

Measurements of boiling performance are shown in Figure 2, which compares the plain SiO₂ surfaces to surfaces featuring hydrophobic or hydrophilic networks. Figure 2a shows the typical heat flux $q''$ vs. superheat $\Delta T=T_w-T_{sat}$ curve. Values of CHF for a plain wafer treated with HF are about 115W/cm² at $\Delta T=27K$, consistent with the 100 W/cm² at $\Delta T=33K$ obtained in [10] for a surface with slightly larger (18º degree) wetting angle, shown in Figure 3. All patterned surfaces exhibit a boiling curve steeper than the plane wafers, with values of CHF ranging from 90 to 190 W/cm², up to 165% of the values of the plain wafer. The highest CHF was reached for a pitch of 100 µm. Patterned surfaces treated with HF exhibit a much higher CHF than untreated patterned surfaces. In the three tested instances, hydrophobic networks exhibit a significantly lower CHF than the hydrophilic networks, sometimes even lower than plain SiO₂ surfaces. Figure 2b shows the HTC as a function of the heat flux. For heat fluxes lower than 50 W/cm², three groups of surfaces can be distinguished by their HTC. Plane surfaces exhibit the lowest HTC, hydrophilic networks show intermediate HTC values, and hydrophobic networks show the maximum HTC values. For heat fluxes higher than 50 W/cm², the HTC of the hydrophilic networks increases to values up to 85kW/m², which is twice the max HTC of the plain SiO₂ surfaces. As a summary, patterning of mixed hydrophilic and hydrophobic areas can improve the CHF and HTC of a plain hydrophilic surface by 65 and 100%, respectively. While surfaces with hydrophilic networks enhance both the CHF and HTC, surfaces with hydrophobic networks seem to only enhance the HTC and might even reduce the CHF. As shown by the comparison in Figure 3, the maximum values obtained in this work are comparable to the maximum HTC and CHF obtained on surfaces covered with a carpet of nanowires [5,6], but slightly lower than sintered wicking surfaces such as [3]. The surfaces studied in this work however are planar while the surface in [3,5,6] can be considered as extended surfaces which promote wicking transport. To emphasize the concept that the enhanced surfaces presented here have more than a unique value for the wetting angle, they are represented by horizontal lines rather than single points in Figure 3.

Explaining the observed trends is challenging because pool boiling is a transient, multiphase phenomenon, visualization is difficult especially for the violent boiling near CHF, and the geometry and wettability of these enhanced surfaces is complex. The following however can be said from the wide theoretical and experimental body of literature summarized in [1,13,14]. First, the enhancement of HTC on patterned surfaces can be explained by the increased availability of active nucleation sites. Indeed, the nucleate boiling theory of Mikic and Rosenhow [15], states that $HTC = q''/\Delta T = K\sqrt{\pi(k_L \sigma C_p)} f D_b^2 Na$, where K is a constant independent of the wetting angle and fluid properties ($k_L \sigma C_p$), $D_b$ is the bubble diameter, $f$ is the bubble release frequency and $Na$ is the density of active nucleation sites. By patterning nucleation sites, we increase the number of nucleation sites, to which HTC is directly proportional. Interestingly, experiments in *Figure 1*c repeatedly show that bubbles start nucleating on the edge hydrophobic patterns. The edges of the lower conductivity patterns also correspond to local maxima of heat flux density, which have been shown to facilitate the onset of nucleate boiling, in experiments with conductive surfaces covered with perforated polymer films [16]. Mikic and Rosenhow's theory might also explain why the HTC of hydrophobic networks is higher than the HTC of hydrophilic networks, since hydrophobic networks offer a larger hydrophobic area, therefore more nucleation sites. Second, the patterns might also constrain the distance between the nucleation sites, which can moderate instabilities and enhance the CHF. Indeed, as stated by Zuber [1,17], the dryout responsible for CHF is caused by Helmoltz instabilities that merge individual bubble columns. On a plain surface the typical pitch λ between the bubble columns is determined by the Taylor instability $\lambda = 2\pi\sqrt{3\sigma/g(\rho_l - \rho_v)} = 27\text{mm}$ [1]. According to the same theory, the critical vapor velocity that triggers Helmoltz instabilities is inversely proportional to $\lambda^{-0.5}$. This analysis concludes to a maximum "practical" CHF value around 110 W/cm$^2$. Let's assume now that the regular patterns investigated in this study have the ability to constrain the wavelength of the instabilities to the pattern pitch $\lambda_p$. In that case, $\lambda_p$ between 200 and 50 μm would multiply the attainable CHF by $(\lambda/\lambda_p)^{0.5}$, a factor between 11 and 23. For hydrophilic networks, the improvement measured experimentally is "only" 1.65, which indicates that other limiting factor might come into play. For instance the kinetic analysis by Schrage [18] determines that for water at atmospheric pressure an absolute theoretical upper bound for the CHF is 16.5 kW/cm$^2$. Third, the observed influence of the HF treatment in increasing CHF (but not HTC) can be explained by the wettability increase of the hydrophilic regions[8] or by the increased difference in wettability between the hydrophobic and hydrophilic regions. Fourth, the fact that hydrophilic networks show a large CHF enhancement, while hydrophobic networks do not show this CHF enhancement can be explained by the droplet boiling experiments in Figure 4, recorded with a high-speed camera. A 3 μL water droplet is gently deposited on two patterned surface heated to an initial temperature of 132ºC, a hydrophilic network on the left, versus a hydrophobic network on the right. At 0.1s on the hydrophilic network, several individual bubbles have nucleated. A very dynamic boiling process occurs then, visible from the strong and fast perturbations on the drop free surface (t=0.77s). Despite the strong boiling, the drop does not move significantly, being hold to the substrate. On the hydrophobic network, the drop shows a completely different behavior: at t=0.1s, the drop does not seem to wet the substrate, as evidenced by the circular shadow under the drop. No individual bubbles are visible, and the drop moves towards the bottom right of the field of view during the evaporation. The total evaporation times of 11 seconds is one order of magnitude larger than the evaporation time on the hydrophilic network. The sliding, absence of individual bubbles, and larger evaporation time suggest the presence of an insulating vapor film between the bubble and the substrate, analog to the Leidenfrost phenomenon. While the transient experiments in Figure 4 are not equivalent to steady state pool boiling experiments, they suggest that hydrophobic networks help nucleation by preventing the early formation of a vapor film, while hydrophobic networks, where vapor bubbles can easily merge, favor early occurrence of CHF. Indeed the hydrophobic networks, unless treated with HF, exhibit a lower CHF than the bare $SiO_2$, probably because they promote the formation of a vapor film. As a final observation, we note that the size of the patterns is compatible with the size of the active nucleation sites predicted by Hsu's theory [19]. It states that for a fluid at saturation temperature, the range of radii of active nucleation sites $\{r_{max}, r_{min}\} = \frac{\delta}{2C_1}\left[1 \pm \sqrt{1 - \frac{4AC_3}{\delta\theta_w}}\right]$. In this equation, $\delta$ is the boundary layer thickness, $C_1 = (1+\cos\varphi)/\sin\varphi$, $C_3 = (1+\cos\varphi)$, $\varphi$ is the wetting angle, $A = 2\sigma T_{sat}/\rho_v h_{Lv}$, and $\theta_w = T_w - T_{sat}$. For the constant heat flux case, assuming free convection and a linear temperature profile in the thermal boundary layer, with a wetting angle $\varphi=110º$ corresponding to the Teflon surface, we obtain a boundary layer thickness of about 400 µm, corresponding to a range of active nucleation sites of 1 to 40 µm at a wall superheat ΔT=5ºC and 0.3-100µm at $\Delta T$=20ºC. This size is compatible with the size of the hydrophobic patterns, which range from about 25 to 100 µm and exhibit offer smaller nucleation sites, as shown in Figure 1c.

In summary, we have demonstrated that surfaces with networks combining hydrophilic and hydrophobic regions significantly enhance the critical heat flux and the heat transfer coefficient during pool boiling. The best enhancement arises with hydrophilic networks featuring hydrophobic islands, which efficiently prevent the formation of an insulating vapor layer.

D.A. is grateful for the help of NSF CAREER grant #0449269, which allowed for his 2008 visit to the Hong Kong laboratory of H.Q. to start this project.

**Figure captions:**

Figure 1:Typical micrographs (a,b) of surfaces with hydrophilic (black) and hydrophobic (grey) zones. The pattern diameter d is the diameter of the inscribed disk. The pattern pitch is p. Surface (a) is hydrophilic with hydrophobic islands, called a hydrophilic network, and noted (+). Surface (b) is hydrophobic with hydrophilic islands, called a hydrophobic network, and noted (-). At low superheat, bubbles typically nucleate at the edge of the hydrophobic patterns (c).

Figure 2: Boiling curves with measured heat flux as a function of the superheat (a) and heat transfer coefficient as a function of heat flux (b), for patterned and plain wafers. Legend shows pitch p in µm, the optional treatment with hydrofluoric acid (HF) and the type of patterned network (+, -, see Figure1).

Figure 3: Critical Heat Flux as a function of wetting angle. Our results are in color, with horizontal lines for cases where the surfaces had regions of mixed wettabilities. Black and grey dots are comparison data on respectively surfaces with controlled wetting properties and a superhydrophilic carpet of nanowires.

Figure 4: The evaporation of a 3 microliter water drop on two patterned surfaces heated at 132ºC, the left surface exhibiting a hydrophilic network and the right surface, a hydrophobic network. For these experiments the patterns are square, with a pitch of 250µm.

# References

[1] V. Carey, Liquid-Vapor Phase Change Phenomena. (Taylor and Francis, 1992).
[2] Y. Nam and Y.S. Jua, Applied Physics letters **93**, 3, (2008).
[3] J.A. Weibel, S.V. Garimella, and M.T. North, international Journal of Heat and Mass Transfer **53**, 4204, (2010).
[4] L.H. Chien and R.L. Webb, Experimental Thermal and Fluid Science **16**, 332, (1998).
[5] C. Li, Z. Wang, P.I. Wang, Y. Peles, N. Koratkar, and G.P. Peterson, Small **4**, 1084, (2008).
[6] R. Chen, M.-C. Lu, V. Srinivasan, Z. Wang, H.H. Cho, and A. Majumdar, Nano Lett **9**, 548, (2009).
[7] S. Kim, H.D. Kim, H. Kim, H.S. Ahn, H. Jo, J. Kim, and M.H. Kim, Experimental Thermal and Fluid Science **34**, 487, (2010).
[8] V.K. Dhir and S.P. Liaw, Journal of Heat Transfer **111**, 739, (1989).
[9] C.H. Wang and V.K. Dhir, Journal of Heat Transfer-Transactions of the ASME **115**, 670, (1993).
[10] C.H. Wang and V.K. Dhir, Journal of Heat Transfer-Transactions of the ASME **115**, 659, (1993).
[11] B.J. Jones, J.P. McHale, and S.V. Garimella, Journal of Heat Transfer-Transactions of the Asme **131**,2009).
[12] P. Zhang and H.H. Qiu, Journal of Micromechanics and Microengineering **18**, 8, (2008).
[13] V.K. Dhir, Annual Review of Fluid Mechanics **30**, 365, (1998).
[14] I.L. Pioro, W. Rohsenow, and S.S. Doerffer, International Journal of Heat and Mass Transfer **47**, 5033, (2004).
[15] B.B. Mikic and W.M. Rohsenow, Journal of Heat Transfer **91**, 245, (1969).
[16] V.A. Antonenko, Journal of Engineering Physics and Thermophysics **54**, 391, (1988).
[17] N. Zuber, Hydrodynamic Aspects of Boiling Heat Transfer, AEC report AECU-4439, 1959.
[18] R.W. Schrage, A theoretical study of interphase mass transfer. (Columbia University Press, New York NY, 1953).
[19] K.-Y. Hsu, ASME Journal of Heat Transfer **84**, 207, (1962).

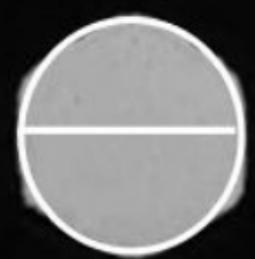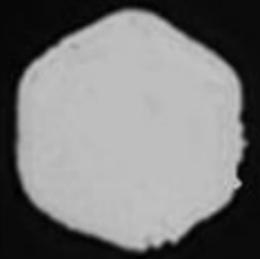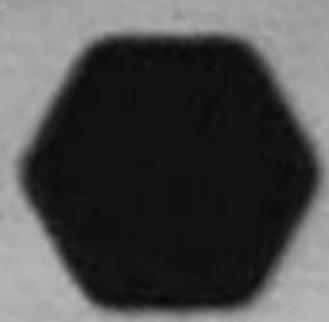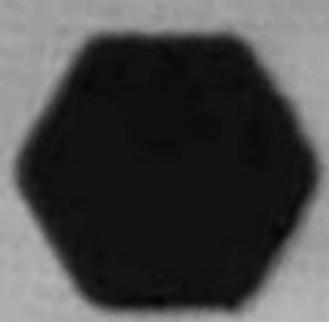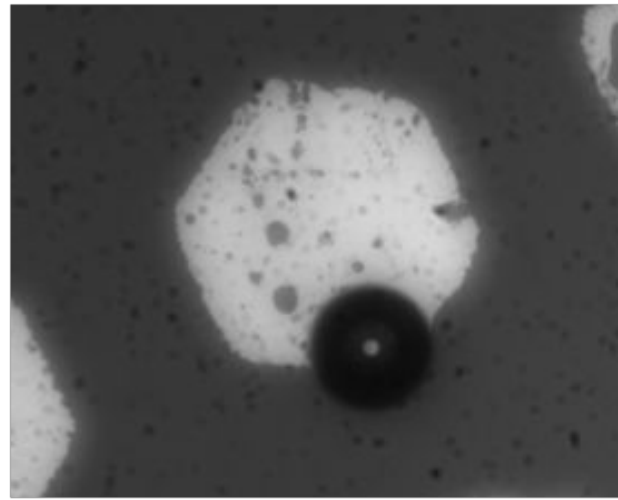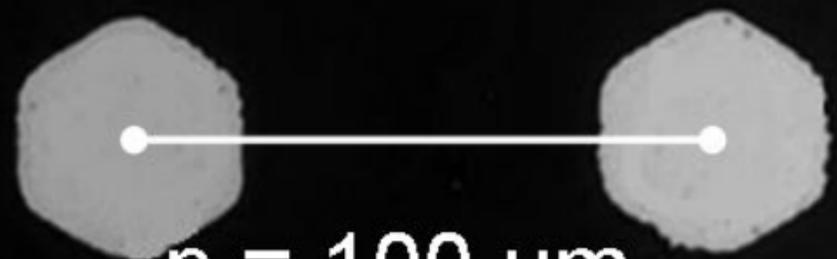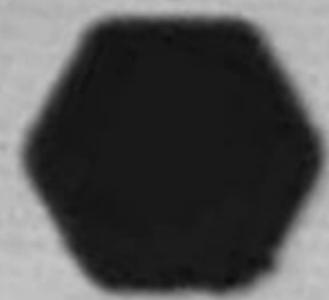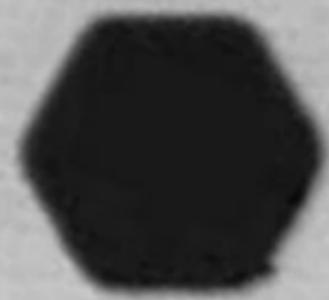

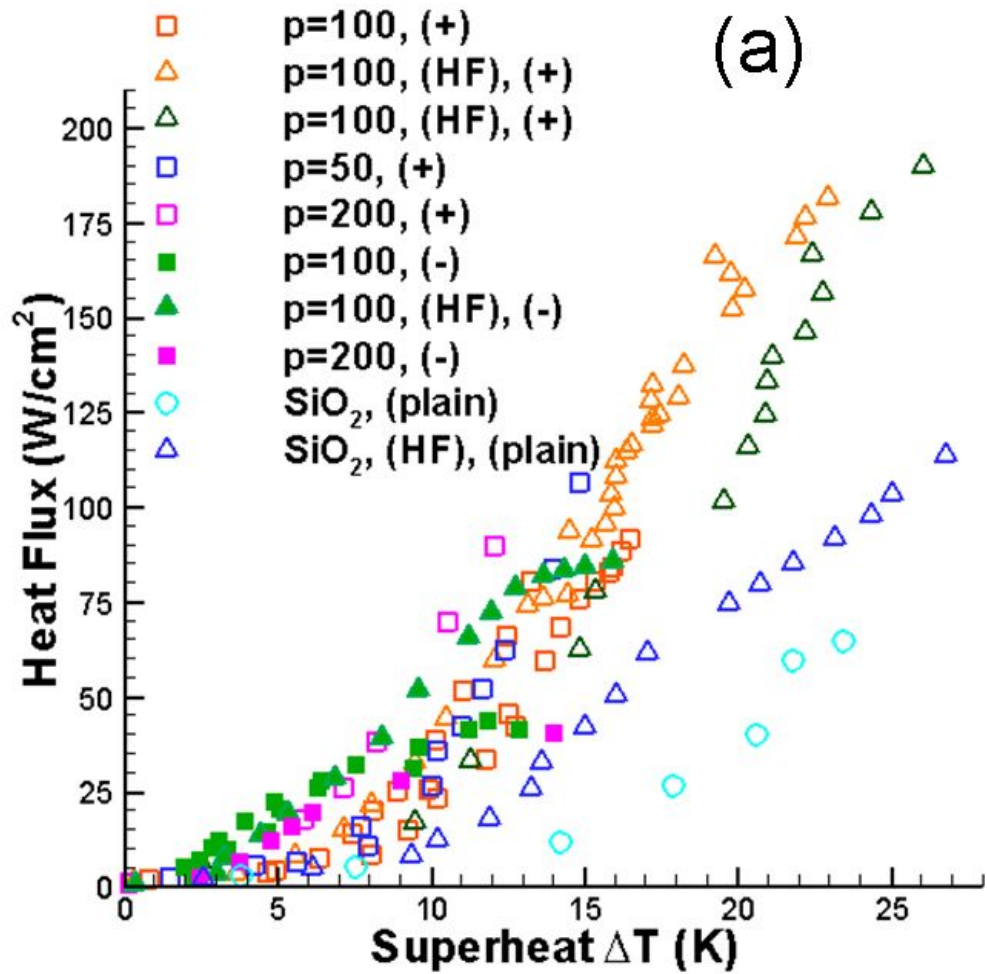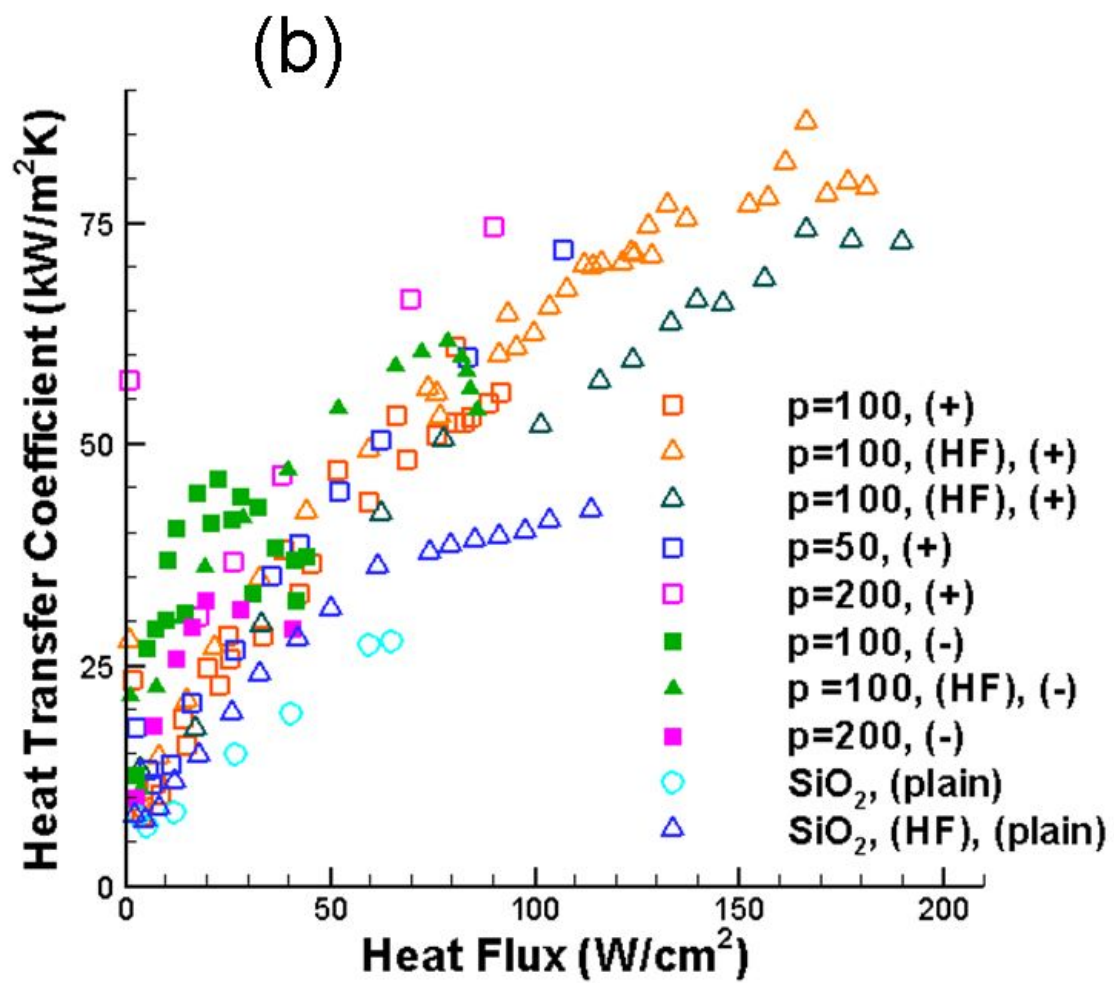

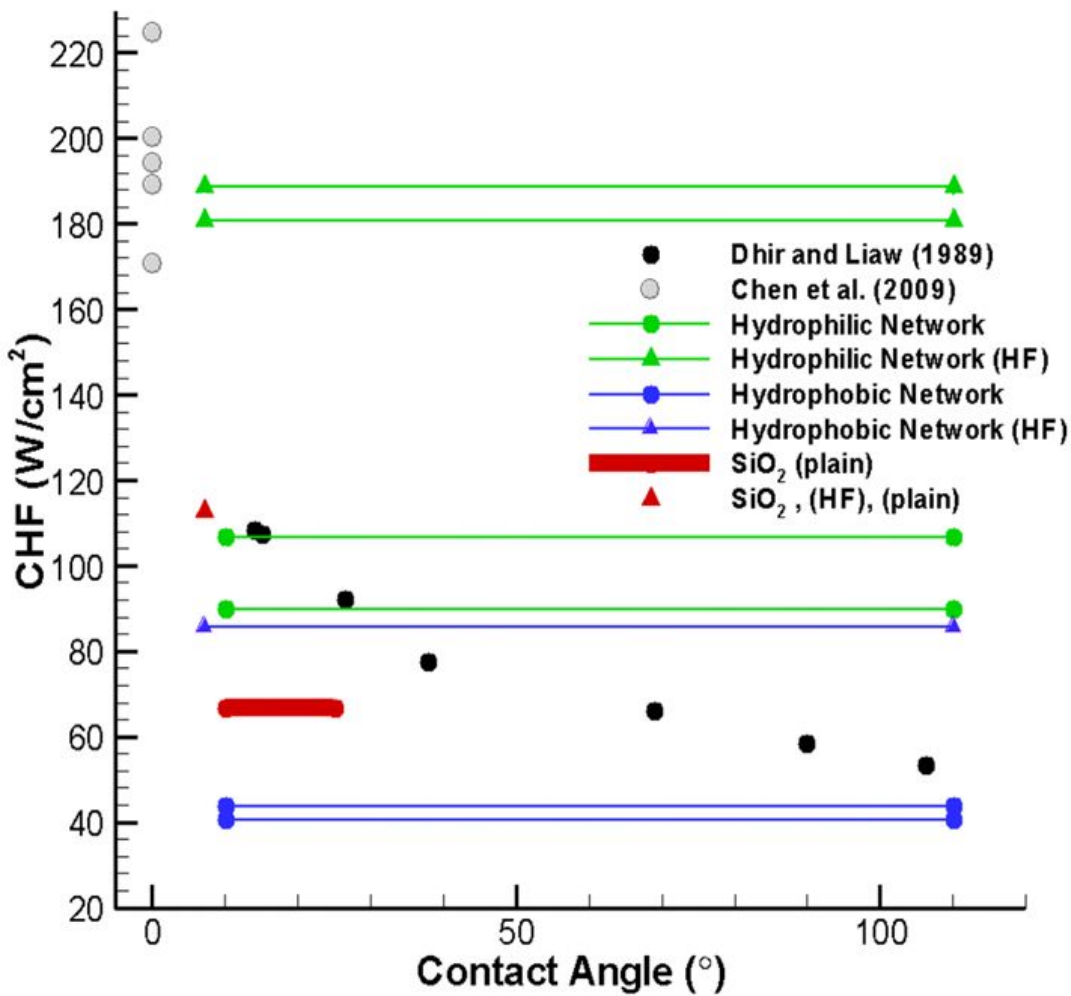

| Hydrophilic Network | Hydrophobic Network |
|---|---|
| 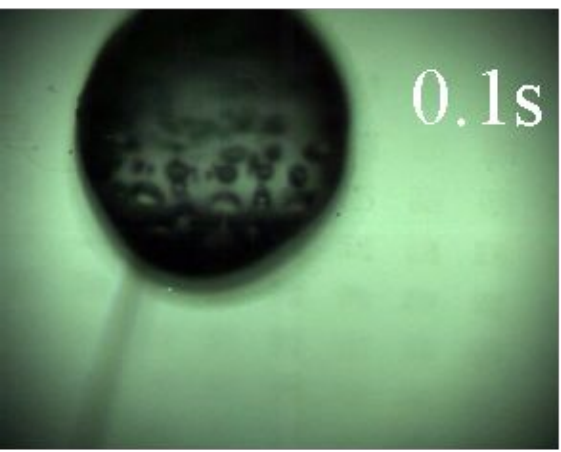 0.1s | 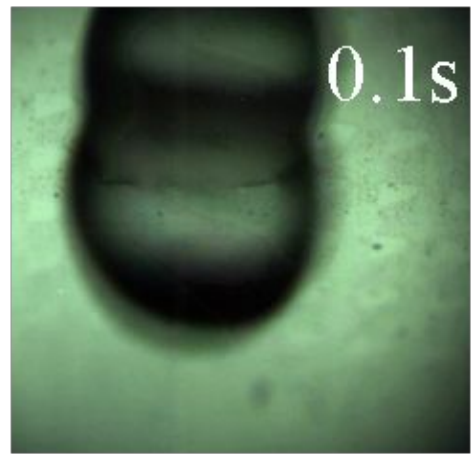 0.1s |
| 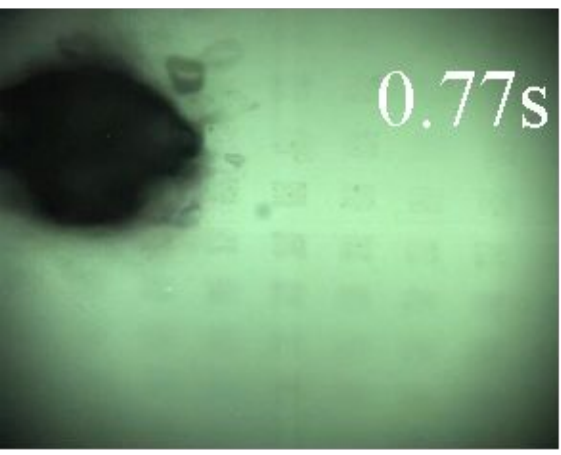 0.77s | 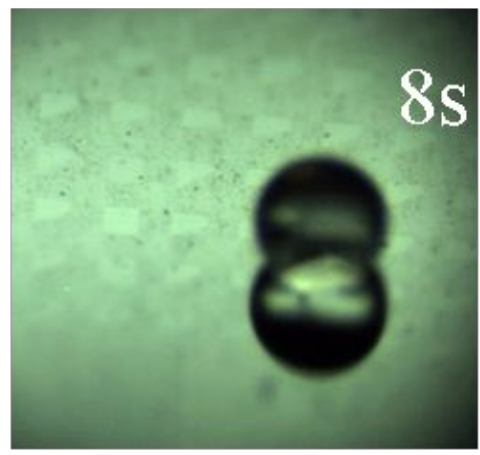 8s |